\documentclass[final,3p,times]{elsarticle}

\usepackage{hyperref}
\usepackage{amssymb}
\usepackage{amsmath}
\usepackage{float}
\usepackage{subcaption}
\journal{Energy Conversion and Management x}

\begin{document}

\begin{frontmatter}




\title{Cost-optimized replacement strategies for water electrolysis systems affected by degradation} 


\author[label1,label2]{Marie Arnold$^{*,}$} 
\author[label2]{Jonathan Brandt$^{*,}$} 
\author[label1]{Geert Tjarks} 
\author[label1]{Anna Vanselow} 
\author[label2]{Richard Hanke-Rauschenbach} 

\affiliation[label1]{organization={EWE GASSPEICHER GmbH},
            city={Oldenburg},
            postcode={26122}, 
            state={Lower Saxony},
            country={Germany}}

\affiliation[label2]{organization={Leibniz Universität Hannover, Institute for Electric Power Systems},
            city={Hanover},
            postcode={30167}, 
            state={Lower Saxony},
            country={Germany}}

\cortext[cor1]{Corresponding authors. Emails: marie.arnold@ewe.de, brandt@ifes.uni-hannover.de}

\begin{abstract}
A key factor in reducing the cost of green hydrogen production projects using water electrolysis systems is to minimize the degradation of the electrolyzer stacks, as this impacts the lifetime of the stacks and therefore the frequency of their replacement. To create a better understanding of the economics of stack degradation, we present a linear optimization approach minimizing the costs of a green hydrogen supply chain including an electrolyzer with degradation modeling. By calculating the levelized cost of hydrogen depending on a variable degradation threshold, the cost optimal time for stack replacement can be identified. We further study how this optimal time of replacement is affected by uncertainties such as the degradation scale, the load-dependency of both degradation and energy demand, and the costs of the electrolyzer. The variation of the identified major uncertainty degradation scale results in a difference of up to 9 years regarding the cost optimal time for stack replacement, respectively lifetime of the stacks. Therefore, a better understanding of the degradation impact is imperative for project cost reductions, which in turn would support a proceeding hydrogen market ramp-up.
\end{abstract}



\begin{keyword}
Green hydrogen \sep Electrolyzer \sep Degradation \sep Renewable energy \sep Techno-economic optimization \sep Stack replacement


\end{keyword}

\end{frontmatter}


\section{Introduction}
Global green hydrogen production by water electrolysis systems, although central to several international governmental hydrogen strategies, is lagging significantly behind ambitions \cite{Odenweller2025}. According to \cite{Odenweller2025}, this is primarily due to an implementation gap entailed by techno-economic constraints, lacking willingness to pay and general early market risks.
In order to achieve these hydrogen strategy goals, it is essential that all relevant cost reduction levers in green hydrogen projects are exploited for the highest possible profitability, concurrently leading to a reduction in investment risks \cite{JohannaReichenbach2023}, \cite{BDEW2023}.\\

\noindent The profitability of a large-scale electrolysis systems is relevantly affected by the degradation of its electrolyzer stacks \cite{Grigoriev2020}.
Significant degradation of the stacks occurs both in constant and in dynamic operation when following a volatile renewable energy profile \cite{Wei2019},\cite{Sayed-Ahmed2024}, which is relevant in the context of decreasing the carbon emissions of the energy sector \cite{InternationalEnergyAgency2023} and the production of green hydrogen.
On the one hand, increasing degradation causes an increase in electricity purchases if certain hydrogen supply contracts include fixed production volumes that must be met \cite{IRENA2020}. 
On the other hand, degradation leads to a reduced stack lifetime \cite{Wei2019},\cite{Alia2019}, which equally affects the frequency of necessary stack replacements during the project operation period. Stack replacements in turn cause additional project costs due to new procurement as well as maintenance periods in which no or reduced hydrogen production can take place \cite{Park2025}. Thus, frequent stack replacements related to degradation lead to additional costs \cite{FfE2025}.\\\noindent \\
Consequently, understanding degradation and its economical influence when operating an electrolyzer is a current research topic being investigated in different approaches:
Zheng et al. 2023 \cite{Zheng2023}, Zhou et al. 2025 \cite{Zhou2025}, Lu et al. 2023 \cite{Lu2023}, Zheng et al. 2024 \cite{Zheng2024}, Luxa et al. 2022 \cite{Luxa2022} and Zhang et al. 2024 \cite{Zhang2024} found that minimizing degradation in multi electrolyzer management systems leads to a more even load distribution across the systems.
Zhang et al. 2022 \cite{Zhang2022} consider optimized electrolyzer operation supplied by renewable energies while modeling degradation afterwards. In contrast, Xu et al. 2023 \cite{Xu2023}, Shin et al. 2023 \cite{Shin2023} and Kuang et al. 2024 \cite{Kuang2024} dynamically adjust electrolyzer operation in response to degradation. Although these studies state the resulting end of life of the electrolyzer, their modeling is not geared to this nor do they mention a universal end of life value.
However, the technological end of life of an electrolyzer is the focus of studies by Krenz et al. 2024 \cite{Krenz2024}, Feng et al. 2017 \cite{Feng2017} and Wang et al. 2023 \cite{Wang2023} as well as the U.S. Department of Energy \cite{USDepartmentofEnergyDoE2022}, each of which define the end of life as a 10\% cell voltage increase. However, it remains unclear if this end of life value is techno-economically reasonable.
Lee et al. 2023 \cite{Lee2023}, Park et al. 2025 \cite{Park2025} and Campbell-Stanway et al. 2025 \cite{Campbell-Stanway2025} examine the impact of degradation on the end of life of electrolysis systems and thus on the technically and economically optimal timing to replace the stacks. 
However, the respective degradation consideration is technology-specific regarding electrolysis technologies and there is no systematic approach addressing any uncertainties. Furthermore, their respective economic analysis does not consider a hydrogen supply chain. This is particularly important for reliable economic evaluation, as consumer demand or hydrogen storage impact the operation of an electrolyzer and thus the degradation of the stacks. This aspect is considered by Köstlbacher et al. 2025 \cite{Kstlbacher2025}, concurrently they did not vary their stack degradation assumption to examine this uncertain parameter. \\\noindent \\
This study presents a comprehensive assessment of the economic impact of degradation on the profitability of a hydrogen supply chain. Based on this, techno-economic optimal stack replacement strategies are systematically explored. 
In Section \ref{Sec: Methodological approach} we introduce the methodical approach, where the economical basis is provided by a linear cost optimization of an integrated onsite hydrogen supply chain including Power Purchase Agreement (PPA) contracting, hydrogen cavern storage and hydrogen demand as described in Section \ref{Sec: System_under_consideration}. This methodological approach is complemented by an electrochemical-based load-dependent and technology-open degradation model, which is introduced in Section \ref{Sec:Increase of the specific energy demand} and Section \ref{Sec: Degradation_Effects}.
In Section \ref{Sec: Result_base_case}, the results of the base case are shown and discussed.
Subsequently, in Section \ref{Sec: Uncertainty_investigations} several uncertainties are examined: the influence of electrolyzer cost variation (Section \ref{Sec: CAPEX_variation}), the unknown impact degradation has on the polarization curve (Section \ref{Sec: Shift_polarization_curve}), the influence of different degradation scales (Section \ref{Sec: Scale_dependency}) and the load-dependency of degradation (Section \ref{Sec: Load_dependency}). Finally, Section \ref{Sec: Overview} concludes with a combination of these uncertainties in a comparative overview.

\section{Methodological approach}\label{Sec: Methodological approach}

\subsection{System under consideration}\label{Sec: System_under_consideration}

\begin{figure}[H]
    \centering \includegraphics[width=0.7\textwidth]
    {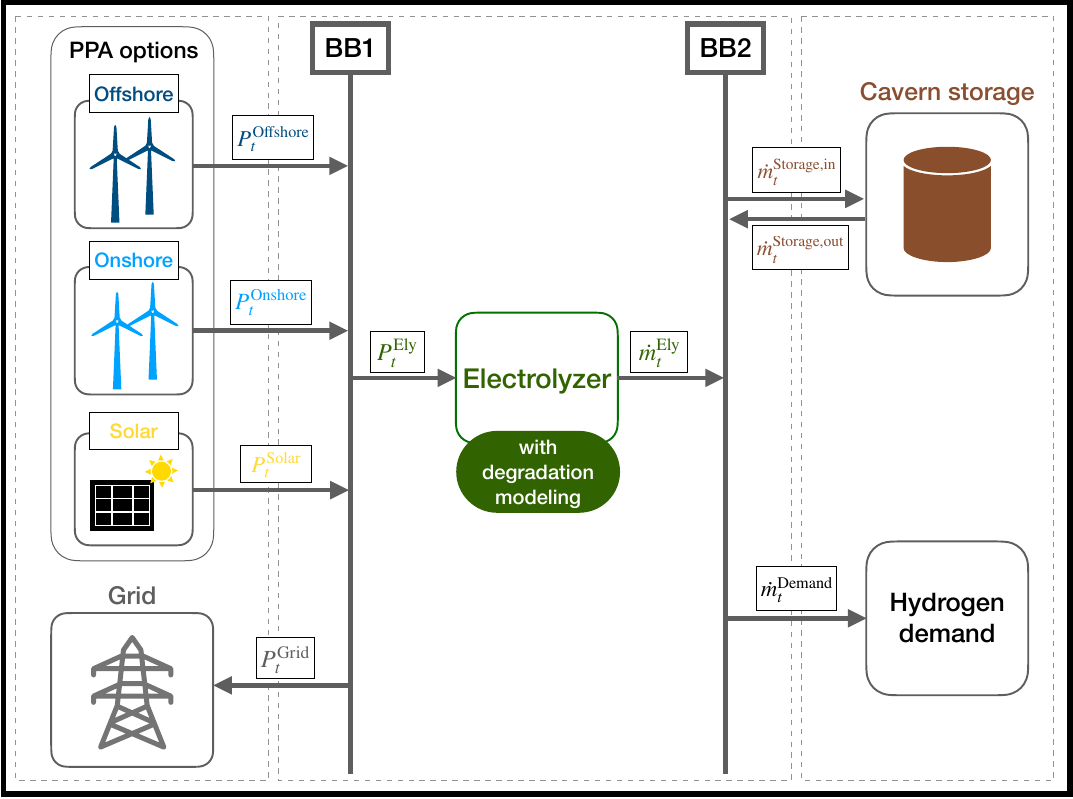}
    \caption{Set-up of the system under consideration. On the left, PPA options are shown including onshore wind, offshore wind and solar energy complemented by the electricity grid. The middle presents the electrolyzer including degradation modeling. On the right, a hydrogen cavern storage is given as well as a consumer characterized by a certain hydrogen demand. BB1 marks the electricity bus bar, BB2 marks the hydrogen bus bar.}
    \label{Fig: System under consideration}
\end{figure}

The methodological base is formed by mathematically minimizing the operational expenditures (OPEX) of one year of the hydrogen supply chain shown in Figure \ref{Fig: System under consideration} to meet a predefined hydrogen demand. The related objective function is shown in Eq. \eqref{Eq: objective_function}. 

\begin{align}\label{Eq: objective_function}
    \text{min}\quad C^\text{PPA} + C^\text{Storage} - R^\text{Surplus}
\end{align}

The OPEX consist of pay-as-produced PPA expenses ($C^\text{PPA}$) plus cavern storage bundle booking expenses ($C^\text{Storage}$) minus surplus sale revenues ($R^\text{Surplus}$), which are based on a constant electricity pricing assumption. 

\noindent The optimization variables of the dispatch problem in Eq. \eqref{Eq: objective_function} are all mass- and energy flows given in Figure \ref{Fig: System under consideration}, as well as the yearly PPA and storage bookings. All constraints of the optimization problem are given in \ref{app1}. The optimization variables and parameters are listed in Table \ref{Tab:Optimization variables} and Table \ref{Tab:Optimization parameters}. The framework for the linear optimization is based on an hourly time scale. For simplicity, the electrolyzer is modeled as a single stack with fixed sizing, omitting any modular system architecture. The predefined hydrogen demand is assumed to be constant, as is typical for industrial applications such as the chemical sector. All economic input parameters are set to present cost values derived from recent publications and converted into euros (2024 value, €$_{2024}$) using the Chemical Engineering Plant Cost Index (CEPCI) \cite{CEPCI}. 

\subsection{Energy demand increase}\label{Sec:Increase of the specific energy demand}

The electrolyzer produces a hydrogen mass flow $\dot{m}^\text{Ely}_t$ by converting electrical power $P_t^\text{Ely}$ depending on its specific energy demand $\epsilon_{j,k}^\text{Ely}$ in each hour $t$. The relation is shown in 
Eq. \eqref{Eq: hydrogen_production}. 

\begin{align}\label{Eq: hydrogen_production}
    \dot{m}^\text{Ely}_t = \frac{P_t^\text{Ely}}{\epsilon_{j,k}^\text{Ely}}
\end{align}

\noindent The energy demand depends on the load range of the electrolyzer, so that operating the electrolyzer at nominal power corresponds to a higher energy demand than the operation at partial load range. This load-dependency of $\epsilon_{j,k}^\text{Ely}$ is modeled by $j \in \text{J}$ points based on the approach from Ref. \cite{Brandt2024}, see \ref{app1} for further details. In addition, the energy demand $\epsilon_{j,k}^\text{Ely}$ depends on the year $k$, because the energy demand increases due to degradation which is modeled on an annual basis. This degradation-induced increase in energy demand is described in Eq. \eqref{Eq: energy_cons_sum}, where $\epsilon_{j,k}^\text{Ely}$ already known from Eq. \eqref{Eq: hydrogen_production} consists of $\epsilon_{j}^\text{Ely, dgr-free}$, which is the energy demand independent of degradation at the electrolyzer's beginning of life (BOL), and $\epsilon_{j,k}^\text{Ely,dgr}$, which describes the degradation surcharge of the respective year $k$. 

\begin{align}\label{Eq: energy_cons_sum}
    \epsilon_{j,k}^\text{Ely} = \epsilon_{j}^\text{Ely, dgr-free} + \epsilon_{j,k}^\text{Ely,dgr} 
\end{align}

\noindent Eq. \eqref{Eq: energy_cons_period} shows, that this degradation surcharge of year $k$ consists of the degradation surcharge $\epsilon_{j,k-1}^\text{Ely, dgr}$ assumed at the beginning of the previous year ($k-1$) and the degradation surcharge delta $\Delta \epsilon_{j,k-1}^\text{Ely,dgr}$ resulting from the previous year ($k-1$). 
Because there is no degradation assumed for the BOL state at the beginning of the first year, it applies that $\epsilon_{j,1}^\text{Ely} = \epsilon_{j}^\text{Ely, dgr-free}$. The calculation of the degradation surcharge delta $\Delta \epsilon_{j,k-1}^\text{Ely,dgr}$ depends on the degradation model under consideration and is further details in the corresponding Section \ref{Sec: Degradation_Effects}.

\begin{align}\label{Eq: energy_cons_period}
    \epsilon_{j,k}^\text{Ely,dgr} = \epsilon_{j,k-1}^\text{Ely, dgr} + \Delta \epsilon_{j,k-1}^\text{Ely,dgr}
\end{align}

\noindent The energy demand increase due to degradation over the years $k$ is limited by a degradation threshold $R$ which corresponds to the end of life (EOL) of the electrolyzer stacks. $R$ is introduced in Eq. \eqref{Eq: R} as the percentage increase of the energy demand at nominal power J in a respective year $k$, $\epsilon_{\text{J},k}^\text{Ely}$, compared to the degradation-free energy demand of the BOL state $\epsilon_\text{J}^\text{Ely,dgr-free}$. In the subsequent analysis, $R$ is varied using the values shown in Section \ref{Sec: Result_base_case}.

\begin{align}\label{Eq: R}
    R = \left( \frac{\epsilon_{\text{J},k}^\text{Ely}}{\epsilon_\text{J}^\text{Ely, dgr-free}} - 1 \right) \cdot 100 \%
\end{align}

\noindent If the increasing energy demand through the years $k$ reaches a predefined threshold value $R$ and thus the EOL of the stacks, the final accounting is made. For this purpose, the averaged levelized cost of hydrogen (LCOH) of the years $k$ from BOL to EOL are calculated, shown in Eq. \eqref{Eq: LCOH}. Each costs of the components of the system under consideration (see Section \ref{Fig: System under consideration}) are included depending on year $k$ and normalized by the total annual hydrogen demand $\sum_{t=1}^T \dot{\text{m}}^\text{Demand}_{t} \cdot \Delta t$. $C_k^\text{PPA}$, $C_k^\text{Storage}$ and $R_k^\text{Surplus}$ are already given by the objective function in Eq. \eqref{Eq: objective_function}. The costs of the electrolyzer, consisting of the costs of the peripherals $C_k^\text{Ely,Peri}$ and the costs of the stacks $C_k^\text{Ely,Stacks}$, are also added. Note that the LCOH with the unit €$_{2024}$/kg represent a standard comparison metric used in the literature \cite{Campbell-Stanway2025} as well as in the hydrogen economy \cite{EvaStede2024}.

\begin{align}\label{Eq: LCOH}
   \text{LCOH}_\text{av} = \sum_{k=1}^{\text{EOL}^\text{Ely,Stacks}} \frac{C_k^\text{PPA} + C_k^\text{Storage}-R_k^\text{Surplus} + C_k^\text{Ely,Peri}+C_k^\text{Ely,Stacks}}{\sum_{t=1}^T \dot{\text{m}}^\text{Demand}_{t} \cdot \Delta t}
\end{align}

\noindent The costs of the peripherals $C_k^\text{Ely,Peri}$ result from equal amortization of the peripheral share of the capital expenditures (CAPEX) over the predefined operating period of the electrolyzer and the fixed annual OPEX, including water costs and maintenance.
The costs of the stacks $C_k^\text{Ely,Stacks}$ result from equal amortization of the stacks share of the CAPEX over the determined life cycle of the stacks from $k = 1$ to $k = \text{EOL}^\text{Ely,Stacks}$. The CAPEX value is the economical uncertainty investigated in the uncertainty analysis in Section \ref{Sec: CAPEX_variation}.

\subsection{Voltage degradation modeling}\label{Sec: Degradation_Effects}

The previously described energy demand increase $\Delta \epsilon_{j,k-1}^\text{Ely,dgr}$, entering Eq. \eqref{Eq: energy_cons_period}, is based on the electrochemically common assumption of a degradation-induced increase of cell voltage over time.
To integrate the voltage increase $\Delta U_{j,k-1}^\text{dgr}$ into the formulation described in Section \ref{Sec:Increase of the specific energy demand}, it must be converted into the energy demand increase $\Delta \epsilon_{j,k-1}^\text{Ely,dgr}$ by means of (with F being the Faraday constant and M$_{\text{H}_2}$ the molar mass of hydrogen):

\begin{align}\label{Eq: Faraday}
    \Delta \epsilon_{j,k-1}^\text{Ely,dgr} = \frac{2\text{F}}{\text{M}_{\text{H}_2}} \cdot \Delta U_{j,k-1}^\text{dgr}
\end{align}

\begin{figure}[H]
    \centering \includegraphics[width=0.5\textwidth]
    {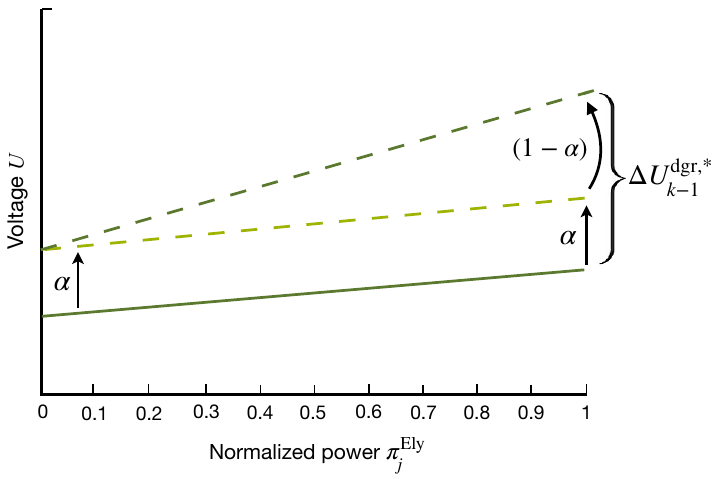}
    \caption{Schematic representation of the voltage increase depending on $\alpha$. The voltage is depending on the normalized power. The dark green continuous line with shows the initial state, the bright green dashed line shows the shift and the dark green dashed line the tilt. Both parts combined result in the total voltage increase in the nominal operation point as presented right.}
    \label{Fig: Alpha}
\end{figure}

\noindent The voltage increase due to degradation affects the polarization curve of the electrolyzer cell in form of a shift, which is independent of the load range, and a tilt, which depends on the load range (\cite{Luxa2022}, \cite{Zhang2024}, \cite{Paciok2017}, \cite{Rakousky2016}). Therefore, the voltage $U$ is modeled depending on the normalized power $\pi^\text{Ely}_j$ as shown in the schematic representation in Figure \ref{Fig: Alpha}. The calculation of the normalized power is given in \ref{app4}. The dark green continuous line describes the initial state in the BOL of the electrolyzer cell. The bright green dashed line shows the shift of the polarization curve and the dark green dashed line the tilt of the polarization curve. Both parts combined result in the total voltage increase $\Delta U_{\text{J},k-1}^\text{dgr}$ at nominal power as presented on the right of Figure \ref{Fig: Alpha}.

\noindent In addition, Figure \ref{Fig: Alpha} shows that the voltage increase $\Delta U_{j,k-1}^\text{dgr}$ depends on $\alpha$, which is used to model the ratio of the mentioned shift and tilt of the polarization curve. The graphical description in Figure \ref{Fig: Alpha} is represented by Eq. \eqref{Eq: load-dependency_Delta_U}, where the first term corresponds to the load-independent shift and the second term corresponds to the load-dependent tilt. Various values between 0 and 1 are assumed for $\alpha$ in order to change the ratio between shift and tilt to investigate this technical uncertainty in the subsequent uncertainty analysis in Section \ref{Sec: Shift_polarization_curve}. A small value assumed for $\alpha$ results in a large tilt share, while a big value assumed for $\alpha$ results in a large shift share.

\begin{align}\label{Eq: load-dependency_Delta_U}
    \Delta U_{j,k-1}^\text{dgr} = \alpha \cdot \Delta U_{k-1}^\text{dgr,*} + \pi_j^\text{Ely} \cdot (1-\alpha) \cdot \Delta U_{k-1}^\text{dgr,*}
\end{align}

\noindent The calculation in Eq. \eqref{Eq: load-dependency_Delta_U} is based on the quantity $\Delta U_{k-1}^\text{dgr,*}$, which represents the degradation voltage increase at nominal power. It is calculated from integrating the degradation rate $\rho^\text{dgr}$ (in µV/h) over time (Eq. \eqref{Eq: sum_rho}. 

\begin{align}\label{Eq: sum_rho}
    \Delta U_{k-1}^\text{dgr,*} = \sum_{t=1}^T \rho^\text{dgr} 
    \left( \pi^{\text{Ely}}_{t,k-1} \right) \cdot \Delta t 
\end{align}

\noindent The degradation rate $\rho^\text{dgr}$ could be either assumed constant or as can be seen in Eq. \eqref{Eq: sum_rho}, being dependent on the load point and thus being dependent on the yearly operating profile following from the optimization described in Section \ref{Sec: System_under_consideration}. Investigating these cases is part of the technical uncertainty analysis in \ref{Sec: Degradation_scenarios}.

\section{Base case definition and initial results}\label{Sec: Result_base_case}

In the first step, a base case is defined by fixing the CAPEX of the electrolyzer, $\alpha$ and the degradation rate $\rho^\text{dgr}$ to average values of the intervals set for these parameters in the subsequent uncertainty analysis in Section \ref{Sec: Uncertainty_investigations}. By this means, the base case applies as a reference for all further investigations.
The results depend on the variation of the degradation threshold $R$ using values shown in Eq. \eqref{Eq: R_variation} and the calculation of the respective average LCOH as described in Eq. \eqref{Eq: LCOH}. 
Furthermore, the degradation rate $\rho^\text{dgr}$ (cf. Eq. \eqref{Eq: sum_rho}) is assumed to be constant.

\begin{align}\label{Eq: R_variation}
    R = [5, 10, 15, 20, 25, 30, 35, 40, 45, 50, 55] \; \%
\end{align}

\begin{figure}[H]
    \centering \includegraphics[width=0.5\textwidth]
    {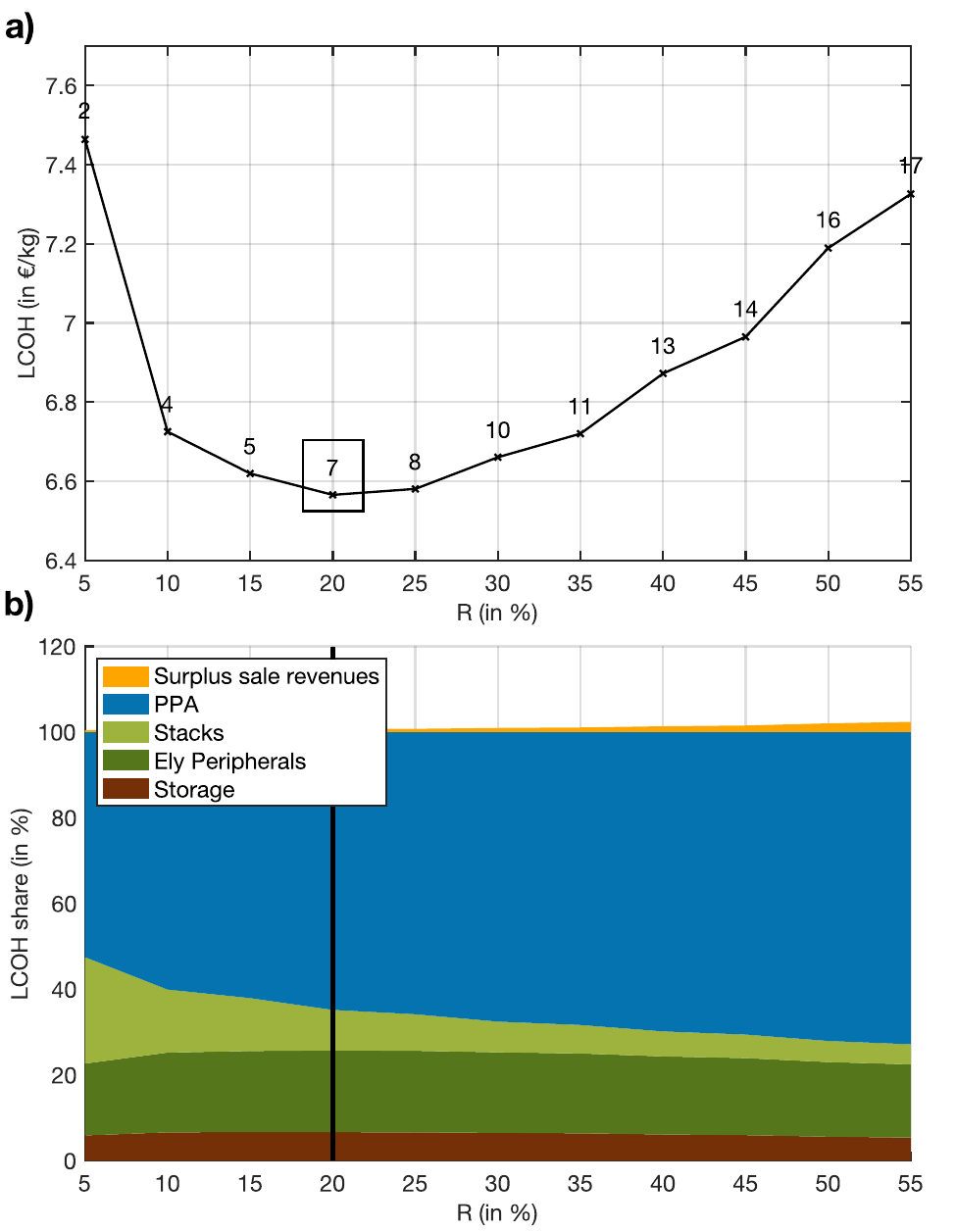}
    \caption{Result of the base case. a) LCOH in €/kg. The replacement period in years is shown for every $R$ above the graph. The minimal LCOH value is marked by a black box. b) Relative cost shares of LCOH of the system’s components. Cost minimal $R$ marked.}
    \label{Fig: Base_case_Result}
\end{figure}

\noindent Figure \ref{Fig: Base_case_Result} a) presents the LCOH in €/kg as a function of $R$. The corresponding replacement periods in years are displayed as numbers above the graph. Figure \ref{Fig: Base_case_Result} b) presents the relative LCOH contributions of the individual system components. 
The minimal LCOH value is marked by a black box in Figure \ref{Fig: Base_case_Result} a) and a black line in Figure \ref{Fig: Base_case_Result} b). In the base case, the minimal LCOH occur at a degradation threshold of $R = 20 \: \%$, which corresponds to a replacement period of 7 years. Figure \ref{Fig: Base_case_Result} b) shows that the higher LCOH for $R$ below the cost minimum result from a higher LCOH share of the stacks. In contrast, the increasing LCOH for $R$ above the cost minimum result from an increasing LCOH share of the PPAs. The minimum of the curve in Figure \ref{Fig: Base_case_Result} a) can therefore be interpreted as the cost optimal time for stack replacement. 

\section{Evaluation of techno-economic uncertainties}\label{Sec: Uncertainty_investigations}

The base case depends on several uncertain techno-economical parameters. The uncertainties investigated in this study are the electrolyzer CAPEX, the technical uncertainty given by the chosen degradation surplus shift/tilt coefficient $\alpha$ (cf. Figure \ref{Fig: Alpha}), as well as the technical uncertainty regarding the scale and load-dependency of the degradation rate $\rho^\text{dgr}$ (cf. Eq. \eqref{Eq: sum_rho}). Each uncertainty variation is compared to the base case discussed in Section \ref{Sec: Result_base_case}.
Subsequently, the dependencies of the outlined uncertainties are collectively and comparatively discussed.

\subsection{CAPEX variation}\label{Sec: CAPEX_variation}

\begin{figure}[H]
    \centering \includegraphics[width=0.5\textwidth]
    {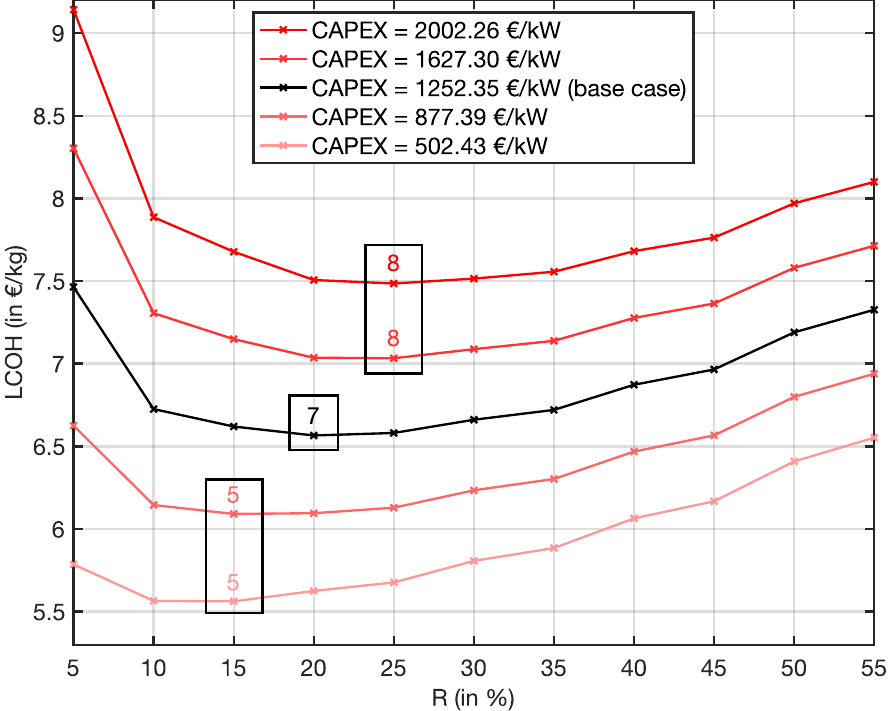}
    \caption{Results of electrolyzer CAPEX variation. LCOH in €/kg over $R$ in $\%$. Different colors show different CAPEX values. Additionally, the replacement period in years is plotted for the minimal LCOH in the color of the respective graph and marked by a black box.}
    \label{Fig: CAPEX_Result}
\end{figure}

The CAPEX of the electrolyzer are varied in order to shift the ratio of the electrolyzer stacks CAPEX share to the PPA expenses, which both have an impact on the LCOH calculation given by Eq. \eqref{Eq: LCOH}. The variation is done based on five CAPEX values given in Eq. \eqref{Eq: CAPEX_variation}. The values are resulting in five graphs shown in Figure \ref{Fig: CAPEX_Result}, whereby the black graph corresponds to the base case already presented in Figure \ref{Fig: Base_case_Result} a). Every graph shows LCOH in €/kg depending on variation of the degradation threshold $R$ in $\%$.

\begin{align}\label{Eq: CAPEX_variation}
    \text{CAPEX} = [502.43,877.39,1252.35,1627.30,2002.26] \; \text{\texteuro/kW}
\end{align}

\noindent The results in Figure \ref{Fig: CAPEX_Result} show a general increase in LCOH as the CAPEX increase. Furthermore, the lowest CAPEX value leads to a replacement period of 5 years and a degradation threshold of 15 $\%$, whereas the highest CAPEX value results in a replacement period of 8 years and a degradation threshold of 25 $\%$. Compared to the base case results, a replacement period of 7 years and a degradation threshold of 20 $\%$, the difference to the results of the lowest CAPEX value is greater than to the results of the highest CAPEX value.
The short replacement period of 5 years for the lowest CAPEX value indicates that PPA expenses dominate stack CAPEX in the LCOH shares, making frequent stack replacements after only 5 years more favorable.
This aspect is going to be further investigated in Section \ref{Sec: Overview}.

\subsection{Variation of degradation surplus shift/tilt coefficient \texorpdfstring{$\alpha$}{alpha}}\label{Sec: Shift_polarization_curve}

The change of the polarization curve described in Section \ref{Sec: Degradation_Effects} is dependent on $\alpha$. The variation of $\alpha$ leads to different shift-tilt-ratios as previously described by Eq. \eqref{Eq: load-dependency_Delta_U} and Figure \ref{Fig: Alpha}.
Eq. \eqref{Eq: Alpha_variation} shows three values assumed for $\alpha$ to analyze the influence of this technical uncertainty on the cost optimal time for stack replacement. 

\begin{align}\label{Eq: Alpha_variation}
    \alpha = [0.075,0.4125,0.75]
\end{align}

\begin{figure}[H]
    \centering \includegraphics[width=0.5\textwidth]
    {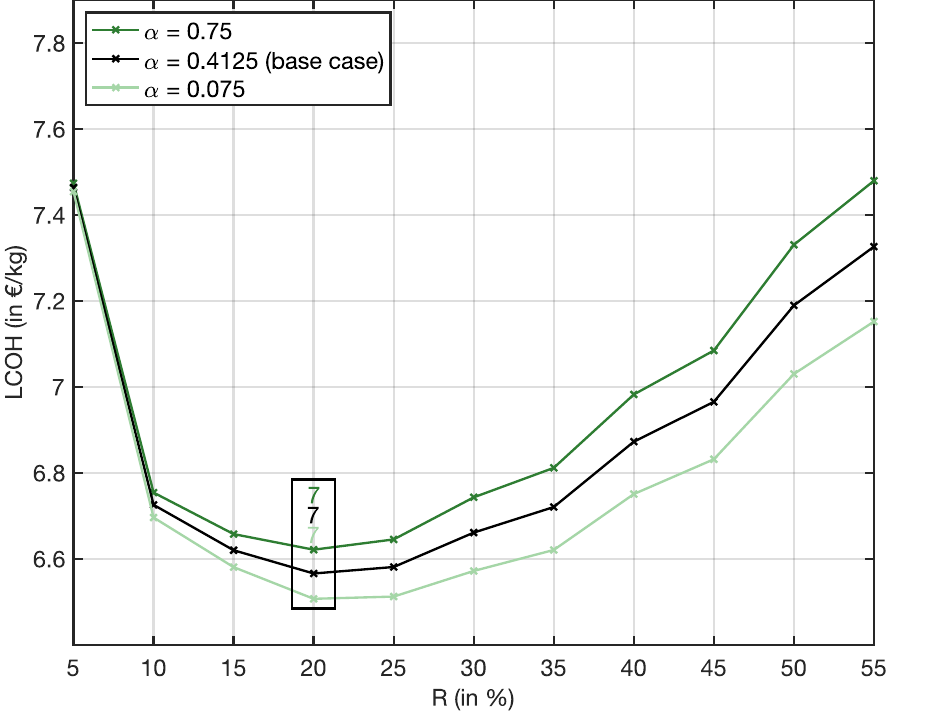}
    \caption{Results of $\alpha$ variation. LCOH in €/kg over $R$ in $\%$. Different colors show different $\alpha$ values. Additionally, the replacement period in years is plotted for the minimal LCOH in the color of the respective graph and marked by a black box.}
    \label{Fig: Alpha_Result}
\end{figure}

\noindent The lowest value $\alpha = 0.075$ represents a larger tilt in relation to the shift, the highest value $\alpha = 0.75$ represents a smaller tilt. $\alpha = 0.4125$ is the average used for the base case results in Section \ref{Sec: Result_base_case}. The results of $\alpha$ variation are shown in Figure \ref{Fig: Alpha_Result} equivalent to the presentation given in Figure \ref{Fig: CAPEX_Result}, whereby the base case corresponds to the black graph here as well.

\noindent Figure \ref{Fig: Alpha_Result} shows a small LCOH delta of approximately $0.12$ €/kg as a consequence of $\alpha$ variation yet no change of the cost optimal time for stack replacement neither regarding replacement period nor degradation threshold. Every minimal LCOH is reached at a replacement period of 7 years and a degradation threshold of 20 $\%$ as it applies to the base case.

\subsection{Variation of degradation scenarios}\label{Sec: Degradation_scenarios}

\begin{figure}[H]
    \centering \includegraphics[width=0.5\textwidth]
    {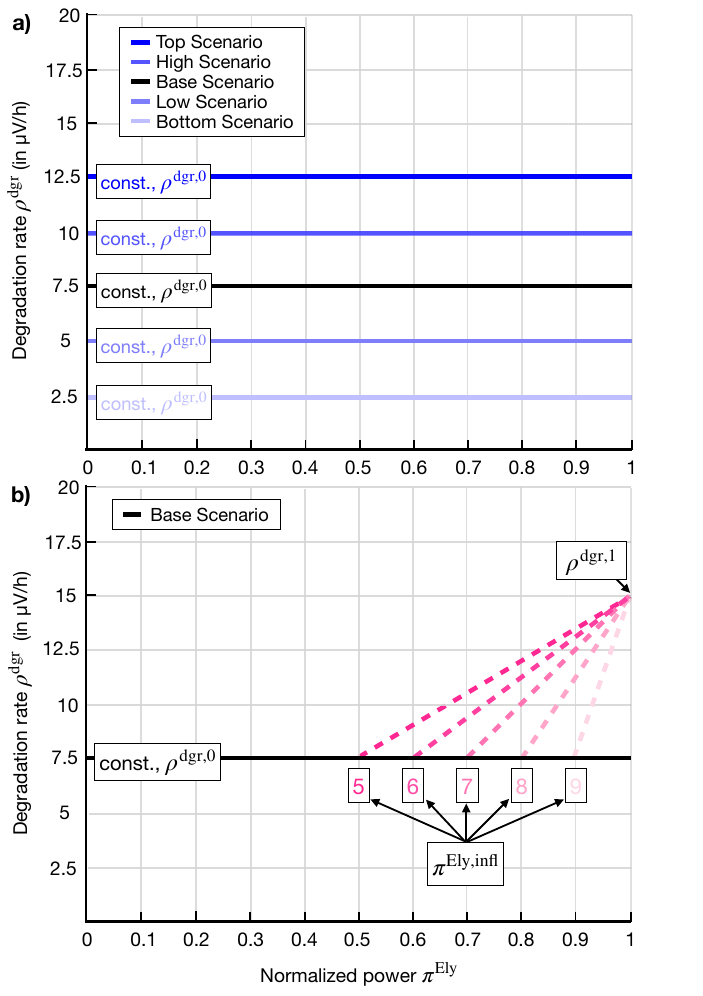}
    \caption{Analyzed degradation scenarios as functions of normalized power. a) Five blue lines show the constant degradation scales. The black line corresponds to the base case. b) Base scenario is given by a continuous black line, while the pink lines show the inflection point options marked by numbers 5 to 9.}
    \label{Fig:Degradation_Scenario_Sketch}
\end{figure}

The base case analyzed in Section \ref{Sec: Result_base_case} is subject to a load-independent, thus constant assumption made for the degradation rate $\rho^\text{dgr}(t)$, which was already mentioned in Section \ref{Sec: Degradation_Effects} and is presented here in Figure \ref{Fig:Degradation_Scenario_Sketch} a) and b) as the black curves. This degradation rate assumption is uncertain according to literature both regarding scale (\cite{Rakousky2016}, \cite{Buttler2018}, \cite{Lettenmeier2016}, \cite{Yu2018}, \cite{Esfandiari2024}, \cite{Fouda-Onana2016}, \cite{Rakousky2017}, \cite{Superchi2023}) as well as load-dependency (\cite{Suermann2019}, \cite{Siracusano2017}). To take this into account, these uncertainties are modeled by degradation scenarios shown in Figure \ref{Fig:Degradation_Scenario_Sketch}. \\

\noindent On the one hand, the scale is considered by a variation of degradation rate values $\rho^\text{dgr,0}$ from 2.5 µV/h to 12.5 µV/h, which remain constant over all load ranges as shown in Figure \ref{Fig:Degradation_Scenario_Sketch} a), where they are labeled accordingly. From small to large, the scales are named bottom, low, high and top, each shown in a different shade of blue, while the base case forms the average and is shown in black. 
On the other hand, the load-dependency in form of a degradation rate increase at high load ranges is also considered by different curve progressions shown in Figure \ref{Fig:Degradation_Scenario_Sketch} b) as dashed lines in shades of pink, that are marked by the numbers 5 to 9. Hereby, 5 describes the inflection point $\pi^{\text{Ely,infl}}$ starting at 50 $\%$ partial load and 9 describes the inflection point at 90 $\%$ partial load. The mathematical description of the degradation rate scenario calculation is given in \ref{app4}. The maximum degradation rate value at nominal load corresponds to a doubling of the base case degradation rate assumption, so that $\rho^\text{dgr,1} = 15$ µV/h for each inflection point.

\subsubsection{Influence of degradation scale}\label{Sec: Scale_dependency}

\begin{figure}[H]
    \centering \includegraphics[width=0.47\textwidth]
    {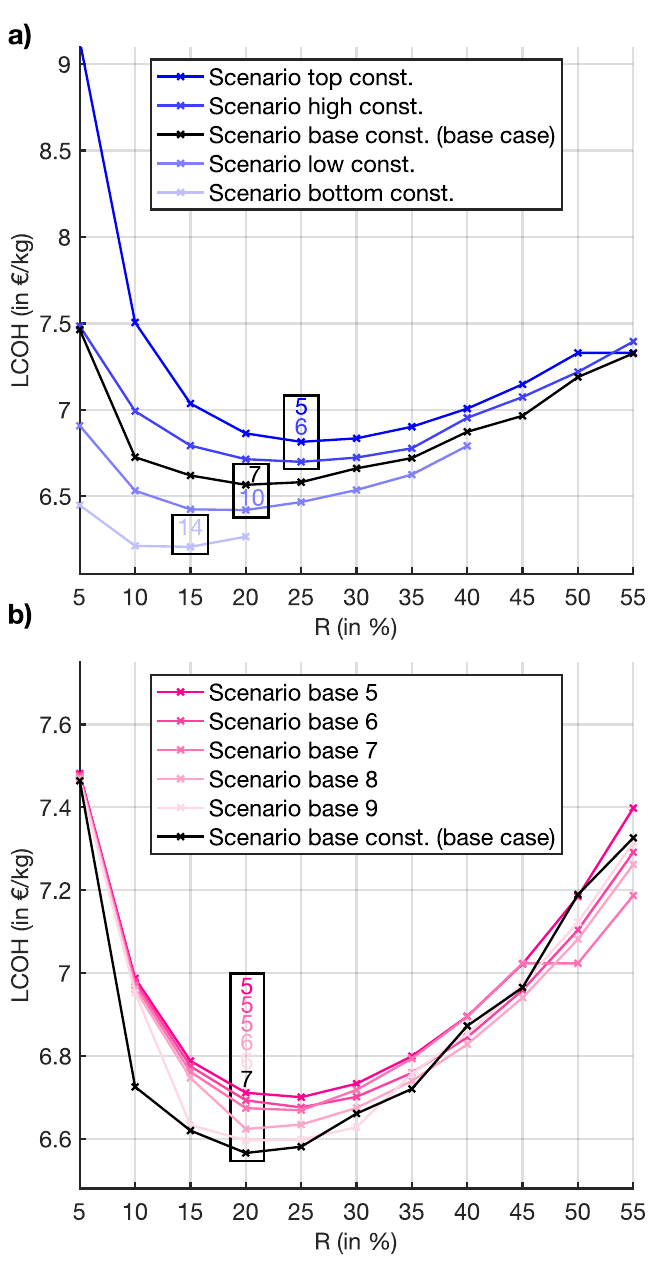}
    \caption{Results of degradation scenario variation. a) Degradation scale variation. LCOH in €/kg over $R$ in $\%$. Different colors show different degradation scales. Additionally, the replacement period in years is plotted for the minimal LCOH in the color of the respective graph and marked by a black box. b) Inflection point variation. LCOH in €/kg over $R$ in $\%$. Different colors show different inflection point options. Additionally, the replacement period in years is plotted for the minimal LCOH in the color of the respective graph and marked by a black box.}
    \label{Fig:Degradation_Result}
\end{figure}

The results of degradation scale variation introduced in Figure \ref{Fig:Degradation_Scenario_Sketch} a) are shown in Figure \ref{Fig:Degradation_Result} a) equivalent to the presentation of the results in previous figures, whereby the base case corresponds to the black graph here as well. The results show increasing LCOH as the degradation scale increases. The minimal LCOH range between 6.21 €/kg and 6.81 €/kg giving a delta of 0.60 €/kg. At these minima, a trend of decreasing cost optimal replacement periods and increasing degradation thresholds is given as the degradation scale scenario increases. More concretely, the replacement period varies between 14 to 5 years and the respective degradation thresholds vary between 15 to 25 $\%$ due to degradation scale increase. Thus, the cost optimal time for stack replacement clearly depends on the chosen scale of the degradation scenario, as a fivefold increase in degradation scale halves the replacement period. With respect to an exemplary twenty-year operating period of the entire electrolysis system, this leads to just one replacement of the stacks for the lowest degradation scenario, bottom const., while for the highest degradation scenario, top const., three stack replacements are cost optimal. 

\subsubsection{Influence of inflection point}\label{Sec: Load_dependency}

The results of the variation of the inflection point in Figure \ref{Fig:Degradation_Scenario_Sketch} b) are shown in Figure \ref{Fig:Degradation_Result} b) equivalent to the presentation of the results in previous figures, whereby the base case corresponds to the black graph here as well. The resulting curves are more similar with showing less effect onto the LCOH than those resulting from the comparison of the degradation scales. More concretely, the minimal LCOH range between 6.59 €/kg and 6.70 €/kg giving a delta of 0.11 €/kg for inflection point variation. Additionally, the replacement period decreases from 6 to 5 years as the inflection point changes to smaller load ranges and the degradation threshold does not change. In summary, the variation of the inflection point therefore has a small influence on the cost optimal time for stack replacement.
This could be because the highest degradation rate is assumed at the highest load range for each inflection point equally, and the highest load range has the most operating hours. Therefore, the energy demand increase due to degradation is similar for each inflection point assumption. Thus, the cost optimal time for stack replacement is sensitive to inflection point variation, but based on the made assumptions the impact is smaller than the degradation scale impact.

\subsection{Multi-parameter variations}\label{Sec: Overview}

In order to see the inter-dependencies of the individual uncertainties and confirm the conclusions drawn, this last results section includes the cost uncertainty given by the CAPEX variation discussed in Section \ref{Sec: CAPEX_variation} as well as the technical uncertainties regarding $\alpha$ discussed in Section \ref{Sec: Shift_polarization_curve} and the degradation scale discussed in Section \ref{Sec: Scale_dependency}. Due to its clear small impact in comparison to degradation scale variation, the degradation load-dependency variation discussed in Section \ref{Sec: Load_dependency} is neglected here. \\

\begin{figure}[H]
  \centering
  \begin{minipage}[t]{0.48\textwidth}
    \centering
    \begin{subfigure}[t]{\textwidth}
      \includegraphics[width=\linewidth]{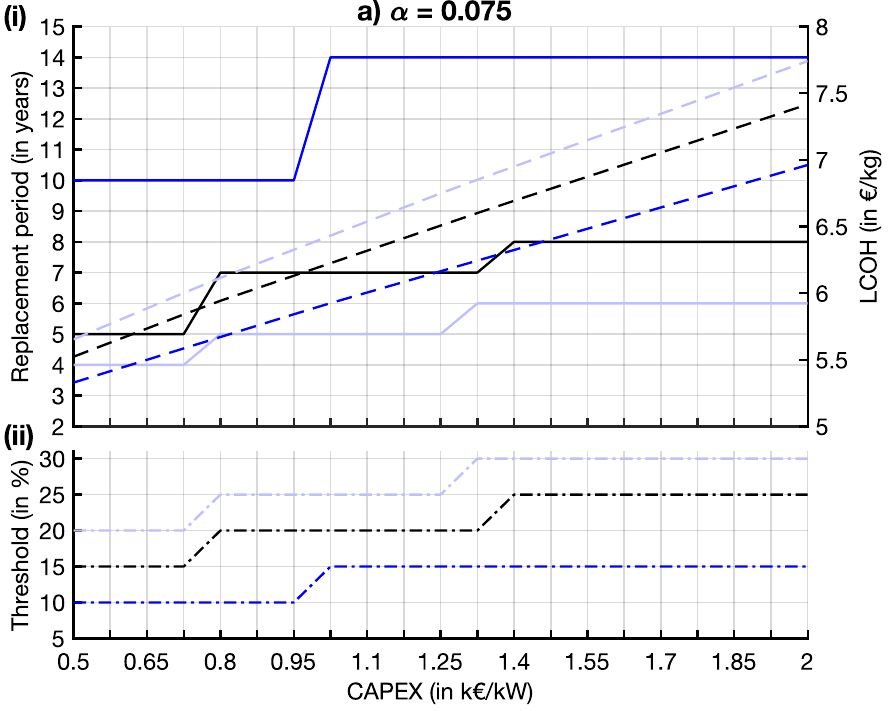}
    \end{subfigure}
    \begin{subfigure}[t]{\textwidth}
      \includegraphics[width=\linewidth]{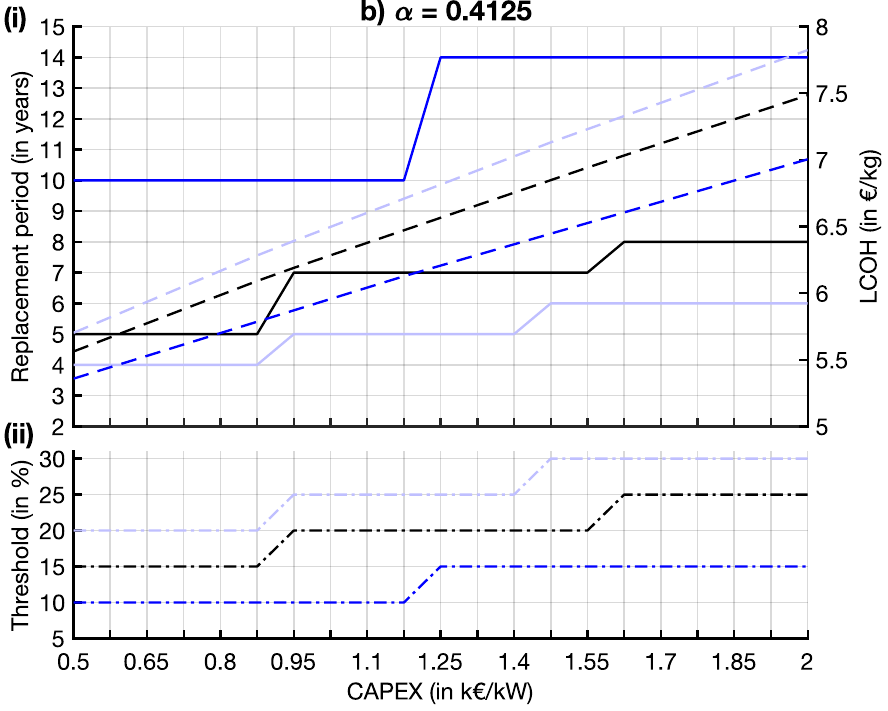}
    \end{subfigure}
    \begin{subfigure}[t]{\textwidth}
      \includegraphics[width=\linewidth]{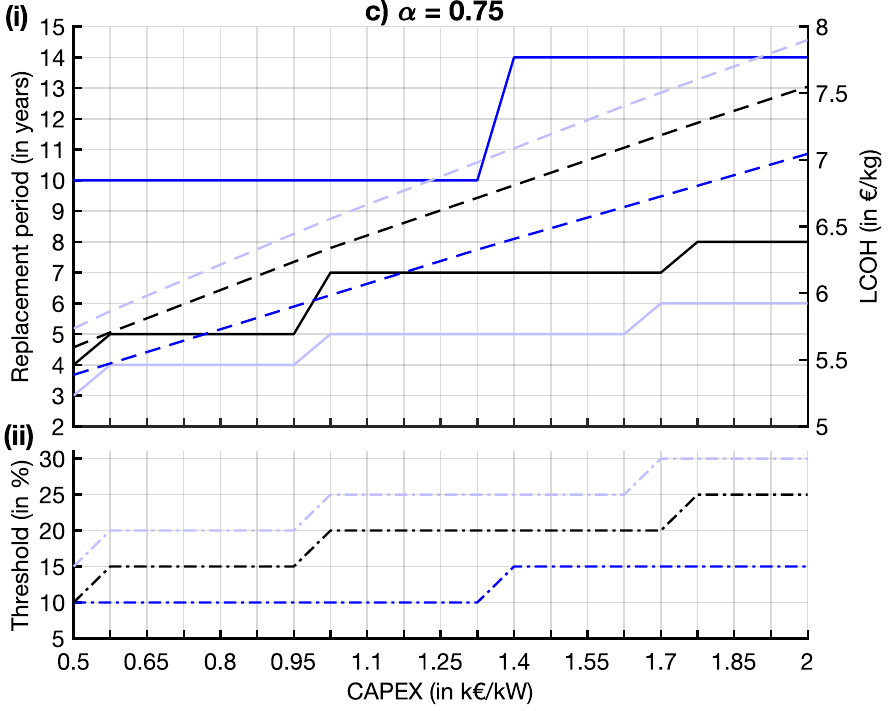}
    \end{subfigure}
    \begin{subfigure}[t]{\textwidth}
    \centering
      \includegraphics[width=0.9\linewidth]{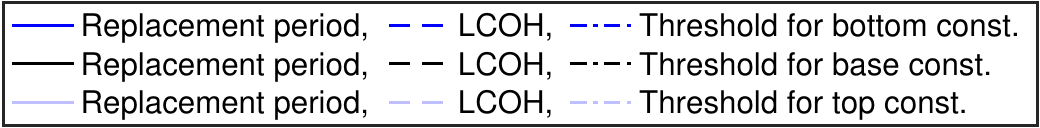}
    \end{subfigure}
  \end{minipage}
  \hfill
  \begin{minipage}[t]{0.48\textwidth}
    \centering
    \begin{subfigure}[t]{\textwidth}
      \includegraphics[width=\linewidth]{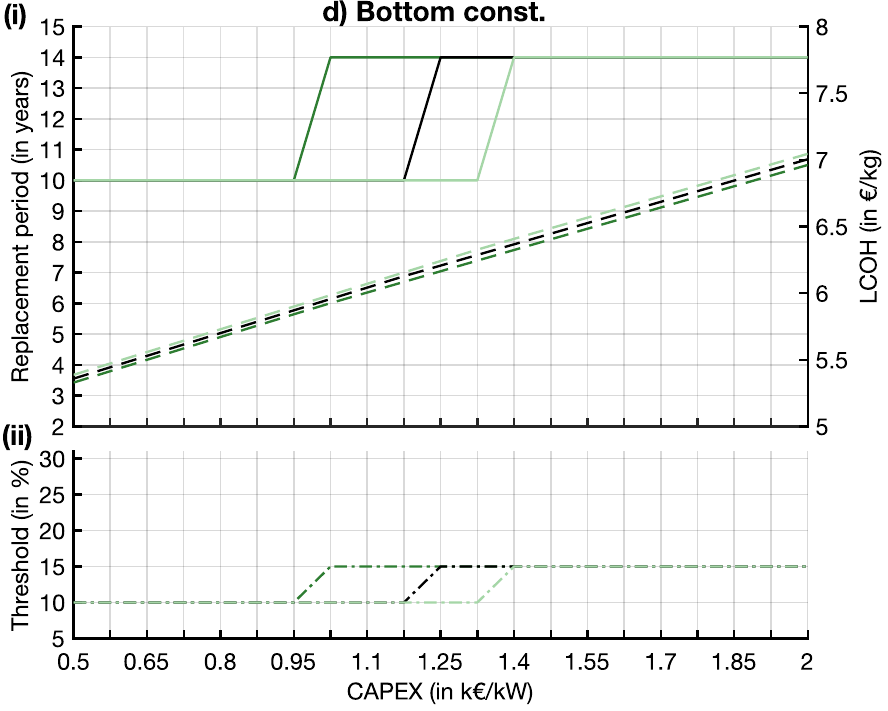}
    \end{subfigure}
    \begin{subfigure}[t]{\textwidth}
      \includegraphics[width=\linewidth]{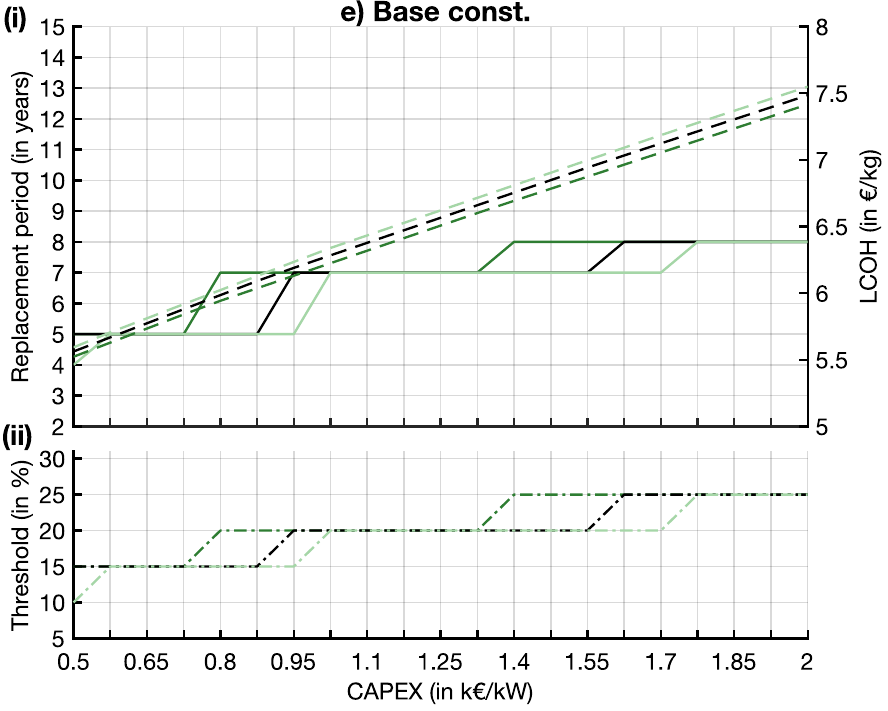}
    \end{subfigure}
    \begin{subfigure}[t]{\textwidth}
      \includegraphics[width=\linewidth]{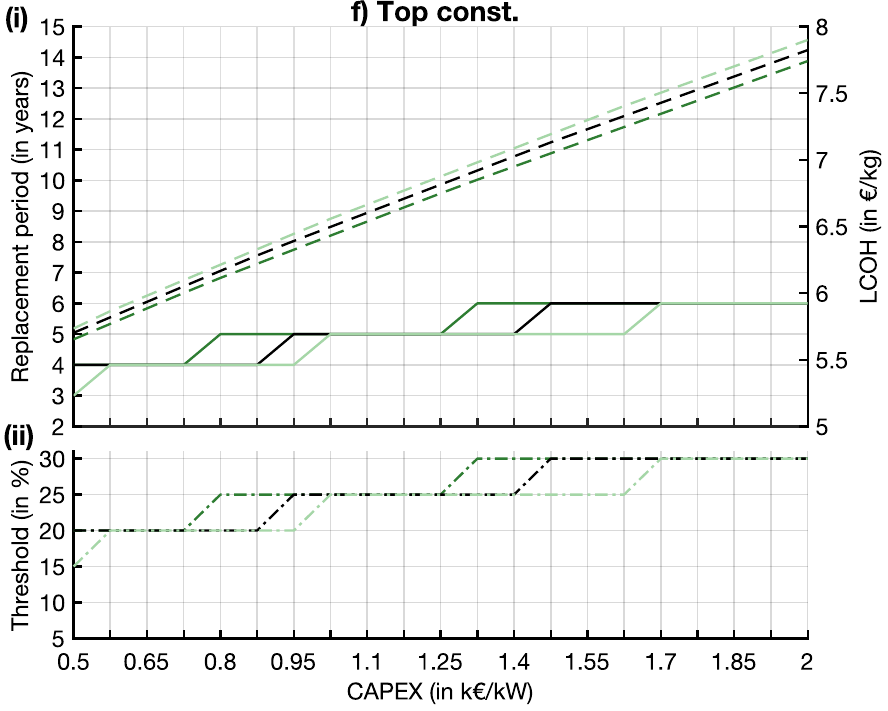}
    \end{subfigure}
    \begin{subfigure}[t]{\textwidth}
    \centering
      \includegraphics[width=0.9\linewidth]{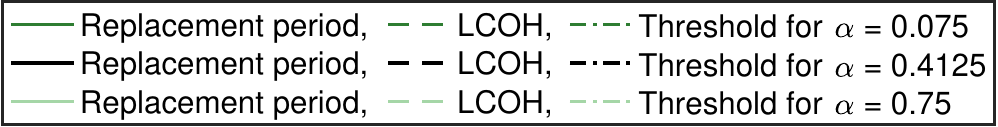}
    \end{subfigure}
  \end{minipage}
  \caption{Results of multi-parameter variations. a)-e) contain two subfigures i) and ii) with the same x-axis presenting the CAPEX variation in k€/kW. i) shows the replacement period in years on the left y-axis and the LCOH in €/kg on the right y-axis, ii) shows the degradation threshold in $\%$ on the y-axis. a), b), c), different graphs show degradation scale variation for different $\alpha$ values locked. d), e), f), different graphs show $\alpha$ variation for different degradation scales locked.}
  \label{Fig: Overview}
\end{figure}

\noindent The CAPEX are varied within the extreme values of the interval given in Eq. \eqref{Eq: CAPEX_variation} and presented on the x-axis of every Subfigure of Figure \ref{Fig: Overview}. Each i) of Subfigures a)-e) shows the cost optimal replacement period with continuous lines on the left y-axis and the minimal LCOH with dashed lines on the right y-axis. Each ii) of Subfigures a)-e) contains the cost optimal degradation threshold on the left y-axis as dashed-dotted lines. Figures \ref{Fig: Overview} a), b) and c) each present an array of curves with the same color range used in Figure \ref{Fig:Degradation_Result} a) to compare degradation scenario top const. with scenario base const. and scenario bottom const. $\alpha$ is locked at each Subfigure for values given in Eq. 12. In contrast, Figures \ref{Fig: Overview} d), e) and f) each present the $\alpha$ variation using the array of curves and the color range from Figure \ref{Fig: Alpha_Result}, while the aforementioned degradation scenarios are locked here. \\

\noindent First, the technical uncertainties of $\alpha$ variation and degradation scale variation are comparatively analyzed. For this purpose, the left column (Figures \ref{Fig: Overview} a), b), c)) containing degradation scale variation is compared with the right column (Figures \ref{Fig: Overview} d), e), f)) containing $\alpha$ variation. The comparison shows that the lines differ less in the right column. Therefore, the impact of $\alpha$ variation on both the cost optimal replacement period and degradation threshold as well as the minimal LCOH is negligible in comparison to the impact of degradation scale variation. 
Second, the impact of degradation scale variation is compared to the impact of the CAPEX variation. For this purpose, Figure \ref{Fig: Overview} b) is considered. For the analysis of both cost optimal replacement period and minimal LCOH dependency, Figure \ref{Fig: Overview} b) i) is considered. For an exemplary CAPEX value of 1400 €/kW, the maximal difference in replacement periods resulting from degradation scale variation is given by a delta of 9 years. In comparison, CAPEX variation leads to a maximal replacement period delta of 4 years, for example in the case of scenario bottom const.
Regarding minimal LCOH, CAPEX variation has a clearly higher impact than degradation scale variation. For the exemplary scenario bottom const., CAPEX variation leads to a maximal delta of the minimal LCOH of 1.64 €/kg whereas degradation scale variation just leads to a maximal delta of the minimal LCOH of 0.82 €/kg. The impact onto the degradation thresholds in Figure \ref{Fig: Overview} b) ii) is similar for both degradation scale variation and CAPEX variation. \\

\noindent Overall, the overview shows that $\alpha$ variation clearly has the smallest impact in comparison to degradation scale variation and CAPEX variation. Regarding the cost optimal time for stack replacement, CAPEX variation and degradation scale variation have a similar impact on the degradation threshold, but degradation scale variation has a higher impact on the replacement period. Furthermore, CAPEX variation has the highest impact on the minimal LCOH. Concurrently, CAPEX are a reliable cost item when planning an electrolysis project, whereas the costs caused by degradation are more unclear. Due to the discussed major impact of the degradation scenarios on the cost optimized replacement period, this should be subject to close attention regarding cost optimized electrolyzer operation.

\newpage

\section{Conclusion}\label{Sec: Conclusion}
The global development of green hydrogen production projects is currently slowing down due to one of the biggest drivers being uncertain project costs. It is therefore mandatory to identify and understand possible cost reduction levers such as additional costs resulting from the degradation of electrolyzer stacks necessitating replacements. \\\noindent \\
In this paper, we found different techno-economic cost optimal replacement strategies and investigated the influence of different technical as well as economical uncertainties. The investigated degradation scale was found to have a major impact on the cost optimal time for stack replacement, as a replacement period difference of up to 9 years results from degradation scale variation. 
Moreover, a minimal LCOH increase of 0.60 €/kg was found for the base case CAPEX value as the degradation scale increases. In contrast, analyzing the influence of inflection point variation for the base case CAPEX value only resulted in a minimal LCOH delta of 0.11 €/kg, with a small replacement period difference of 1 year and no change in the degradation threshold. This makes it negligible in comparison to degradation scale variation. Furthermore, the variation of the degradation surplus shift/tilt coefficient $\alpha$, illustrating the technical uncertainty of the polarization curve shift in this study, also showed negligible influence on the cost optimal time for stack replacement in comparison to degradation scale variation. The economical uncertainty of the CAPEX variation showed a smaller impact regarding replacement period than the degradation scale variation, but a higher impact regarding minimal LCOH. However, due to its higher predictability, the overall impact was rated lower. \\\noindent \\ 
These results can help to understand the extent of the impact of degradation on the economics and replacement times and thus to avoid unnecessarily frequent stack replacements. The degradation scale was identified as the most important uncertainty having a comparatively high LCOH impact combined with a difficult predictability. Therefore, the degradation rate, which forms the basis of the degradation scale variation in this paper, should be subject of further electrochemical and technical studies to improve reliability regarding the planning and operation of green hydrogen production projects.  \\\noindent \\
We explicitly underline that the conclusions drawn in this study are based on the assumptions made. One critical aspect is the yearly modeled degradation surcharge, which is therefore not dynamically minimized. Thus, the cost optimization is not able to consider degradation costs, which could be subject of further investigations. Moreover, start-stop-scenarios involving both cold starts and warm starts of the electrolyzer did not occur in this study and were therefore outside of the scope. However, they could lead to additional degradation effects. Therefore, examining this could also be subject of further investigations. \\

\noindent Due to these limitations and the complexity of the subject, this study alone cannot be expected to display every specification of degradation and examine its sensitivities. Nevertheless, the findings should improve understanding of the cost relevance of degradation effects on green hydrogen projects as well as how additional stack replacement costs can be minimized or even avoided.

\section*{CRediT authorship contribution statement}

\textbf{Marie Arnold:} Writing - original draft, writing - review \& editing, conceptualization, data curation, formal analysis, investigation, methodology, software, visualization. \textbf{Jonathan Brandt:} Writing - review \& editing, conceptualization, methodology, supervision, formal analysis. \textbf{Geert Tjarks:} Writing - review \& editing, conceptualization, supervision. \textbf{Anna Vanselow:} Writing - review \& editing, conceptualization. \textbf{Richard Hanke-Rauschenbach:} Writing - review \& editing, conceptualization, methodology, supervision. 

\section*{Declaration of competing interest}

The authors declare that they have no known competing financial interest or personal relationships that could have appeared to influence the work reported in this paper.

\section*{Data and code availability}

The code and data for this study are publicly available at Zenodo \cite{Zenodo}: \url{https://doi.org/10.5281/zenodo.16877893}. 

\section*{Acknowledgements}

The authors gratefully acknowledge funding by the German Federal Ministry for Economic Affairs and Energy (BMWE) within project SyNerGy-H2 (grant number 03EI6143). The results presented were achieved by computations carried out on the cluster system at Leibniz Universität Hannover, Germany.

\appendix
\section{Mathematical description of optimization problem}\label{app1}

\noindent The objective function in Eq. \eqref{Eq.: obj} aims to minimize the total annual cost of the system by optimizing the design and operation of the considered components (compare with Eq. (1) in main document).

\begin{align}\label{Eq.: obj}
    \min \quad C^{\text{PPA}} + C^{\text{Storage}} - R^{\text{Surplus}}
\end{align}

\noindent The total PPA expenses $C^{\text{PPA}}$ are calculated as a double summation over all time steps $t$ and PPA options, shown in Eq. \eqref{Eq.: PPA costs}. Thereby, the individual PPA costs are calculated as the product of the rated nominal power of the respective PPA option $P^{\text{PPA}}$, the specific power purchase price $\text{p}^{\text{PPA}}$, the capacity factor $\text{f}_t^{\text{PPA}}$, and the length of time step $\Delta t$.

\begin{align}\label{Eq.: PPA costs}
    C^{\text{PPA}} = \sum_{t=1}^{\text{T}} \sum_{\text{PPA}} \text{p}^{\text{PPA}} \cdot P^{\text{PPA}} \cdot \text{f}_t^{\text{PPA}} \cdot \Delta t
\end{align}

\noindent The costs of the hydrogen storage $C^{\text{Storage}}$ 
are calculated as follows by adding the booking costs resulting from multiplying the capacity fee $\text{p}^{\text{Storage,cap}}$ with the booked storage capacity $m^{\text{Storage,max}}$ to the operational costs resulting from multiplying the usage fee $\text{p}^{\text{Storage,turn}}$ and the sum of the time-dependent mass flow rate $\dot{m}^{\text{Storage,in}}_{t}$ multiplied by the length of time step $\Delta t$.

\begin{align}
    C^{\text{Storage}} = \text{p}^{\text{Storage,cap}} \cdot m^{\text{Storage,max}} + \text{p}^{\text{Storage,turn}} \cdot \sum_{t=1}^T \dot{m}^{\text{Storage,in}}_{t} \cdot \Delta t
\end{align}

\noindent The surplus revenues $R^{\text{Surplus}}$ are calculated by multiplying the sum of the surplus power $P^{\text{Surplus}}_{t}$ by the grid electricity price $\text{p}^{\text{Grid}}$ and the length of time step $\Delta t$, shown in Eq. \eqref{Eq: Surplus}.

\begin{align}\label{Eq: Surplus}
    R^{\text{Surplus}} = \sum_{t=1}^T P^{\text{Surplus}}_{t} \cdot \text{p}^{\text{Grid}} \cdot \Delta t
\end{align}

\noindent The following equality constraints define the technical operation of the system under consideration shown in Figure 1 in the main document. The respective optimization variables are listed in Table \ref{Tab:Optimization variables} and the parameters in Table \ref{Tab:Optimization parameters}.
\begin{align}
  0 &= \dot{m}^{\text{Ely}}_{t} - \left( \dot{m}_{t}^{\text{Storage,in}} - \dot{m}_{t}^{\text{Storage,out}} \right) - \dot{m}_{t}^{\text{Demand}} && \forall t \in \{1,2,3,\dots,\text{T}\} \\
  0 &= P_t^{\text{Onshore}} + P_t^{\text{Offshore}} + P_t^{\text{Solar}} - P_t^{\text{Ely}} - P_t^{\text{Grid}} && \forall t \in \{1,2,3,\dots,\text{T}\} \\
  m_t^{\text{Storage}} &= m_{t-1}^{\text{Storage}} + \left( \dot{m}_t^{\text{Storage,in}} - \dot{m}_t^{\text{Storage,out}} \right) \cdot \Delta t && \forall t \in \{2,3,\dots,\text{T}\} \\
  m_1^{\text{Storage}} &= m_T^{\text{Storage}} + \left( \dot{m}_1^{\text{Storage,in}} - \dot{m}_1^{\text{Storage,out}} \right) \cdot \Delta t && \forall t \in \{1,2,3,\dots,T\}
\end{align}

\noindent In the following, the inequality constraints of the optimization problem are shown.

\begin{align}
  P_t^{\text{Onshore}} &\leq P^{\text{PPA,Onshore}} \cdot f_t^{\text{PPA,Onshore}} && \forall t \in \{1,2,3,\dots,\text{T}\} \\
  P_t^{\text{Offshore}} &\leq P^{\text{PPA,Offshore}} \cdot f_t^{\text{PPA,Offshore}} && \forall t \in \{1,2,3,\dots,\text{T}\} \\
  P_t^{\text{Solar}} &\leq P^{\text{PPA,Solar}} \cdot f_t^{\text{PPA,Solar}} && \forall t \in \{1,2,3,\dots,\text{T}\} \\
  P_t^{\text{Grid}} &\geq 0 && \forall t \in \{1,2,3,\dots,\text{T}\} \\
  0 &\leq m_t^{\text{Storage}} \leq m^{\text{Storage,max}} && \forall t \in \{1,2,3,\dots,\text{T}\} \\
  0 &\leq \dot{m}_t^{\text{Storage,in}} \leq \dot{m}^{\text{Storage,in,max}} && \forall t \in \{1,2,3,\dots,\text{T}\} \\
  0 &\leq \dot{m}_t^{\text{Storage,out}} \leq \dot{m}^{\text{Storage,out,max}} && \forall t \in \{1,2,3,\dots,\text{T}\}
\end{align}

\noindent The load-dependency of the energy demand of the electrolyzer, shown in Eq. (2) in the main document, is integrated into the optimization problem by the linearization method used in \cite{Brandt2024}. The respective constraints constructing a convex search space are shown in the following, with the y-axis intersect $b^{\text{lin}}_j$ of the respective linear constraint $j$.

\begin{align}
    \dot{m}^{\text{Ely}}_{t} - a^{\text{lin}}_j \cdot P_t^{\text{Ely}} - b^{\text{lin}}_j &\leq 0 && \forall t \in \{1,2,3,\dots,\text{T}\},\ \forall j \in \{1,2,3,\dots,\text{J}{-}1\}
\end{align}

\begin{align}
    a^{\text{lin}}_j &= \frac{j+1}{\epsilon^{\text{Ely}}_{j+1}} - \frac{j}{\epsilon^{\text{Ely}}_{j}}
\end{align}

\begin{align}
    b^{\text{lin}}_j &= \frac{\frac{j}{\text{J}-1} \cdot \text{P}^{\text{Ely,Nom}}}{\epsilon^{\text{Ely}}_j} -
    a^{\text{lin}}_j \cdot \frac{j}{\text{J}-1} \cdot \text{P}^{\text{Ely,Nom}}
\end{align}

\noindent The number of linearization steps used in this study is 37. To reduce the search space and thereby the optimization time, a lower bound is defined and shown in Eq.~\ref{Eq:lower_bound}.

\begin{align}\label{Eq:lower_bound}
    \text{P}^{\text{Ely,Nom}} - \dot{m}^{\text{Ely}}_{t} \cdot \epsilon^{\text{Ely,Nom}} &\leq 0 && \forall t \in \{1,2,3,\dots,\text{T}\}
\end{align}

\begin{table}[H]
\centering
\caption{Optimization variables}
\renewcommand{\arraystretch}{1.5} 
\begin{tabular}{ |p{2.5cm}|p{10cm}| } 
\hline
Variable & Description \\
\hline
$P_t^{\text{Onshore}}$         & Onshore wind power at time step $t$ \\
$P_t^{\text{Offshore}}$        & Offshore wind power at time step $t$ \\ 
$P_t^{\text{Solar}}$           & Solar power at time step $t$\\ 
$P^{\text{PPA,Onshore}}$       & Booked onshore PPA for one year \\ 
$P^{\text{PPA,Offshore}}$      & Booked offshore PPA for one year \\ 
$P^{\text{PPA,Solar}}$         & Booked solar PPA for one year\\ 
$P_t^{\text{Ely}}$             & Electrolyzer power consumption at time step $t$ \\ 
$P_t^{\text{Grid}}$            & Surplus power at time step $t$ \\ 
$\dot{m}_t^{\text{Ely}}$             & Hydrogen produced by the electrolyzer at time step $t$  \\ 
$\dot{m}_t^{\text{Storage,in}}$     & Hydrogen stored at time step $t$ \\ 
$\dot{m}_t^{\text{Storage,out}}$    & Hydrogen provided by the storage at time step $t$ \\ 
$m_t^{\text{Storage}}$         & Stored hydrogen mass at time step $t$ \\ 
$m^{\text{Storage,max}}$       & Booked storage capacity for one year \\ 
\hline
\end{tabular}
\label{Tab:Optimization variables}
\end{table}

\begin{table}[H]
\centering
\caption{Optimization parameters}
\renewcommand{\arraystretch}{1.5} 
\begin{tabular}{ |p{2.5cm}|p{10cm}|} 
\hline
Parameter & Description \\
\hline
$a^{\text{lin}}_j$        & Gradient of linearized characteristic curve of the electrolyzer between linearization steps $j$ and $j+1$ \\ 
$b^{\text{lin}}_j$        & Y-axis intersect of the linearized characteristic curve of the electrolyzer between linearization steps $j$ and $j+1$ \\ 
$\text{f}_t^{\text{Onshore}}$    & Onshore capacity factor at time step $t$\\ 
$\text{f}_t^{\text{Offshore}}$   & Offshore capacity factor at time step $t$ \\ 
$\text{f}_t^{\text{Solar}}$      & Solar capacity factor at time step $t$ \\
$\text{P}^{\text{Ely,Nom}}$      & Nominal electrolyzer power \\ 
$\epsilon^{\text{Ely}}_{j}$     & Specific energy demand of electrolyzer at $\frac{j}{J-1} \cdot 100\%$ of nominal power\\ 
$\epsilon^{\text{Ely,Nom}}$    & Specific energy demand of electrolyzer at nominal power\\ 
$\dot{\text{m}}^{\text{Demand}}_{t}$  & Predefined hydrogen demand at time step $t$\\
$\Delta t$                & Length of time step in hours\\ 
\hline
\end{tabular}
\label{Tab:Optimization parameters}
\end{table}

\section{Calculation of the electrolyzer costs}\label{app2}

\noindent As shown in Eq. (6) in the main document, for calculating the LCOH, the electrolyzer costs are considered in addition to the costs resulting from the optimization if the degradation threshold $R$ is exceeded. The electrolyzer costs are split into the peripherals costs $C_k^{\text{Ely,Peri}}$ and the stack costs $C_k^{\text{Ely,Stacks}}$. $C_k^{\text{Ely,Peri}}$ is defined as the sum of the peripherals CAPEX costs $C_k^{\text{Ely,Peri,CAPEX}}$ and the OPEX costs $\text{C}_k^{\text{Ely,OPEX}}$, shown in Eq. \eqref{Eq:Peri}.

\begin{align}\label{Eq:Peri}
    C_k^{\text{Ely,Peri}} = C_k^{\text{Ely,Peri,CAPEX}} + \text{C}_k^{\text{Ely,OPEX}}
\end{align}

\noindent The calculation of $C_k^{\text{Ely,Peri,CAPEX}}$ is defined in Eq. \eqref{Eq:Peri_CAPEX} as the multiplication of the nominal power of the electrolyzer $\text{P}^{\text{Ely,Nom}}$ with the costs of the electrolyzer $c^{\text{Ely,CAPEX}}$, the share of these costs of the peripherals $\text{S}^{\text{Ely,Peri}}$ and the annuity factor of the peripherals $\text{A}^{\text{Ely,Peri}}$.

\begin{align}\label{Eq:Peri_CAPEX}
    C^{\text{Ely,Peri,CAPEX}}_k = \text{P}^{\text{Ely,Nom}} \cdot c^{\text{Ely,CAPEX}} \cdot \text{s}^{\text{Ely,Peri}} \cdot \text{A}^{\text{Ely,Peri}}
\end{align}

\noindent $\text{A}^{\text{Ely,Peri}}$ is defined as shown in Eq. \eqref{Eq:Peri_Annuity} with the interest rate representing the real weighted average cost of capital $\text{r}_{\text{in}}^{\text{Ely}}$ and the depreciation time of the electrolyzers' peripherals $\text{t}^{\text{Ely,Peri}}_{\text{dep}}$.

\begin{align}\label{Eq:Peri_Annuity}
    \text{A}^{\text{Ely,Peri}} = \frac{\text{r}_{\text{in}}^{\text{Ely}} \cdot \left(1 + \text{r}_{\text{in}}^{\text{Ely}}\right)^{\text{t}^{\text{Ely,Peri}}_{\text{dep}}}}{\left(1 + \text{r}_{\text{in}}^{\text{Ely}}\right)^{\text{t}^{\text{Ely,Peri}}_{\text{dep}}} - 1}
\end{align}

\noindent The OPEX costs $\text{C}^{\text{Ely,OPEX}}$ add up from the maintenance costs of the electrolyzer $\text{c}^{\text{Ely,OPEX}}$ and the water costs of the electrolyzer $\text{c}^{\text{Water}}$ multiplied with the specific water consumption $\text{w}^{\text{Ely}}$, the yearly demand $\dot{\text{m}}^{\text{Demand}}_{t}$ and the length of time step $\Delta t$.

\begin{align}\label{Eq:Peri_OPEX}
    \text{C}^{\text{Ely,OPEX}} = \text{P}^{\text{Ely,Nom}} \cdot \text{c}^{\text{Ely,OPEX}} + \text{c}^{\text{Water}} \cdot \text{w}^{\text{Ely}} \cdot \sum_{t=1}^\text{T} \dot{\text{m}}^{\text{Demand}}_{t} \cdot \Delta t 
\end{align}

\noindent The calculation of $C_k^{\text{Ely,Stacks}}$ is defined in Eq. \eqref{Eq:Stacks_CAPEX} as the multiplication of the nominal power of the electrolyzer $\text{P}^{\text{Ely,Nom}}$ with the costs of the electrolyzer $\text{c}^{\text{Ely,CAPEX}}$, the share of these costs of the stacks $\text{s}^{\text{Ely,Stacks}}$ and the annuity factor of the stacks $A^{\text{Ely,Stacks}}$.

\begin{align}\label{Eq:Stacks_CAPEX}
    C^{\text{Ely,Stacks}}_k = \text{P}^{\text{Ely,Nom}} \cdot c^{\text{Ely,CAPEX}} \cdot \text{s}^{\text{Ely,Stacks}} \cdot A^{\text{Ely,Stacks}}
\end{align}

\noindent $A^{\text{Ely,Stacks}}$ is defined as shown in Eq. \eqref{Eq:Stacks_Annuity} with the interest rate representing the real weighted average cost of capital $\text{r}_{\text{in}}^{\text{Ely}}$ and the depreciation time of the electrolyzer stacks $t^{\text{Ely,Stacks}}_{\text{dep}}$.

\begin{align}\label{Eq:Stacks_Annuity}
    A^{\text{Ely,Stacks}} = \frac{\text{r}_{\text{in}}^{\text{Ely}} \cdot \left(1 + \text{r}_{\text{in}}^{\text{Ely}}\right)^{t^{\text{Ely,Stacks}}_{\text{dep}}}}{\left(1 + \text{r}_{\text{in}}^{\text{Ely}}\right)^{t^{\text{Ely,Stacks}}_{\text{dep}}} - 1}
\end{align}

\section{Technical and economical parameter assumptions}\label{app3}

\noindent The data for the capacity factors of the renewables used in Eq. \eqref{Eq.: PPA costs} was taken from \cite{RenewablesNinja}, whereby the solar data is based on \cite{Pfenninger2016} and the wind data is based on \cite{Staffell2016}. Weather year 2023 was chosen in this study as well as three renewable power production sites located in Northern Germany. The exact configuration is given in Table \ref{renewableninjatable}.

\begin{table}[H]
\centering
\caption{Renewables configuration}
\renewcommand{\arraystretch}{1.5} 
\begin{tabular}{ |p{3cm}|p{3cm}|p{3.5cm}|p{2.5cm}|} 
\hline
Parameter & Wind onshore & Wind offshore & Solar \\
\hline
Local time  & Europe/Berlin  & Europe/Berlin & Europe/Berlin  \\ 

Location & Dietrichsfeld  & Riffgat & Oldenburg  \\ 

Electricity & kW & kW & kW \\ 

Latitude & 53.5288° & 53.6903° & 53.1756°  \\ 

Longitude & 7.4704° & 6.4811° & 8.1719°  \\ 

Dataset & Merra2 & Merra2 & Merra2  \\ 

Capacity & 1 & 1 & 1  \\ 

Tilt/Azimuth & / & / & 35°/180°  \\ 

Height & 135 m & 90 m & / \\ 

Turbine & Enercon E126 3500 & Siemens SWT 3.6 120 & /  \\ 
\hline
\end{tabular}
\label{renewableninjatable}
\end{table}

\begin{table}[H]
\centering
\caption{Technical and economical parameter assumptions}
\renewcommand{\arraystretch}{1.5} 
\begin{tabular}{ |p{2cm}|p{1.5cm}|p{3.3cm}|p{1.3cm}|p{1.8cm}|p{1.3cm}|} 
\hline
Component & Parameter & Description & Value & Unit & Reference \\
\hline
Wind onshore & $\text{p}^{\text{PPA,Onshore}}$& Pay-as-produced price & 0.0729 & €$_{2024}$/kWh & \cite{Brandt2025}\\ 
\hline
Wind offshore & $\text{p}^{\text{PPA,Offshore}}$& Pay-as-produced price & 0.0883 & €$_{2024}$/kWh & \cite{Brandt2025}\\ 
\hline
Solar & $\text{p}^{\text{PPA,Solar}}$ & Pay-as-produced price & 0.0555 & €$_{2024}$/kWh & \cite{Brandt2025} \\ 
\hline
Grid & $\text{p}^{\text{Grid}}$ & Electricity price & 0.1976 & €$_{2024}$/kWh & \cite{Brandt2025} \\ 
\hline
 Electrolyzer & $\text{P}^{\text{Ely,Nom}}$  &  Nominal power  & 300 000 & kW & Own assumption \\ 
 & $c^{\text{Ely,CAPEX}}$ & CAPEX interval  & [502.43, 2002.26] & €$_{2024}$/kW & \cite{Holst2021}, \cite{Wasike-Schalling2023} \\ 
  & $\text{c}^{\text{Ely,OPEX}}$ & Maintenance OPEX fix  & 23.45 & €$_{2024}$/(kW$\cdot$a) & \cite{Holst2021} \\ 
 & $\text{s}^{\text{Ely,Peri}}$ &  Cost share peripherals & 75 & \% & \cite{Holst2021} \\ 
 & $\text{s}^{\text{Ely,Stacks}}$ &  Cost share stacks & 25 & \% & \cite{Holst2021} \\ 
 & $\text{t}^{\text{Ely,Peri}}_{\text{dep}}$ & Depreciation time peripherals & 20 & a & \cite{UdoLubenau2022} \\ 
 & $\text{r}_{\text{in}}^{\text{Ely}}$ & Interest rate  & 7 & \% & \cite{Brandt2025} \\ 
 & $\epsilon^{\text{Ely,Nom}}$ & Specific energy demand at nominal load  & 52.5 & kWh/kg & \cite{Brandt2025} \\ 
 &  & Decrease of specific energy demand  & 1 & \% per 10 \% load reduction & \cite{Brandt2025} \\
 & $\text{w}^{\text{Ely}}$ & Specific water consumption  & 14 & kgH$_2$O/kgH$_2$ & \cite{Brandt2025}\\ 
 & $\text{c}^{\text{Water}}$ & Water costs  & 3.725 & €$_{2024}$/m$^3$H$_2$O & \cite{Brandt2025}\\ 
\hline 
Hydrogen cavern storage & $\text{p}^{\text{Storage,cap}}$ & Capacity fee  & 12.75 & $\frac{\text{€}_{2024}}{\text{kg} \cdot \text{a}}$ & \cite{Brandt2025} \\
 & $\text{p}^{\text{Storage,turn}}$ & Usage fee  & 0.36 & $\frac{\text{€}_{2024}}{\text{kg} \cdot \text{a}}$ & \cite{Brandt2025}\\ 
 \hline
 Hydrogen demand & $\dot{\text{m}}^{\text{Demand}}_{t}$ &  Amount  & 3200* & kg/h & Own assumption\\ 
\hline
\end{tabular}
\end{table}

\noindent *The combination of the predefined nominal power of the electrolyzer and the predefined hydrogen demand lead to full load hours of approximately 5000 to 6000 hours depending on the respective degradation impact.

\section{Degradation rate calculations}\label{app4}

\noindent The basis for the degradation rate calculations is given by the normalized power $\pi^\text{Ely}$, which is used in Section 2.3. and Section 4.3. in the main document. The normalized power of each time step $t$, $\pi^\text{Ely}_t$, results from dividing the electrolyzer power $P^\text{Ely}_t$ by the nominal power of the electrolyzer $P^\text{Ely,Nom}$ as shown in Eq. \eqref{Eq: normalized power}.

\begin{align}\label{Eq: normalized power}
   \pi^\text{Ely}_t = \frac{P^\text{Ely}_t}{P^\text{Ely,Nom}} 
\end{align}

\noindent Eq. \eqref{Eq: rho_load_dependent} gives the mathematical description of the curve progression of the degradation rate $\rho^{\text{dgr}}(\pi^\text{Ely})$ described in Section 2.3. as well as in Section 4.3. in the main document, where it is shown in Figure 6 b). $\rho^\text{dgr,0}$ describes the base scale of the degradation rate scenario. $\rho^\text{dgr,1}$ describes the degradation rate of the scenario at nominal load, which corresponds to a doubling of the respective $\rho^\text{dgr,0}$. $\pi^\text{Ely,infl}=[0.5,0.9]$ refers to the respective inflection points mentioned in Figure 6 b) in the main document. Note that for the degradation scale scenarios shown in Figure 6 a) in the main document, $\rho^{\text{dgr}}(\pi^\text{Ely})$ is always independent of the load range and therefore $\rho^{\text{dgr}}(\pi^\text{Ely})$ = $\rho^\text{dgr,0}$ for all load ranges.

\begin{align}\label{Eq: rho_load_dependent}
    \rho^{\text{dgr}}(\pi^\text{Ely} ) =
    \begin{cases}
    \:\rho^{\text{dgr},0} & \text{for } \pi^\text{Ely} \leq \pi^\text{Ely,infl} \\
    \:\rho^{\text{dgr},0} + \dfrac{\rho^{\text{dgr},1} - \rho^{\text{dgr},0}}{1 - \pi^\text{Ely,infl}} \left( \pi^\text{Ely}  - \pi^\text{Ely,infl} \right) & \text{for } \pi^\text{Ely} > \pi^\text{Ely,infl}
    \end{cases}
\end{align}

\section{Software}\label{app5}
\noindent The optimization problem was implemented in Matlab \cite{MATLAB}. Gurobi \cite{gurobi} was used as the mathematical solver for all optimizations run in this study.

\bibliographystyle{elsarticle-num}
\bibliography{refs}

\end{document}